# A Continuum Theory of Elastic Semiconductors with Consideration of Mobile Charge Inertia

Jiashi Yang (jyang1@unl.edu)
Department of Mechanical and Materials Engineering
University of Nebraska-Lincoln, Lincoln, NE 68588-0526, USA

**Abstract**
A set of nonlinear equations for the macroscopic theory of elastic semiconductors is derived which generalizes the seminal work of Tiersten by including the inertia of the mobile charge carriers. The equations obtained can describe carrier plasma waves and their interactions with elastic waves. Another generalization in this paper is that the internal energy densities of the mobile charges are allowed to depend on temperature in addition to charge densities. The equations are in SI units.

## 1. Introduction

In the nonlinear theory of elastic semiconductors in [1], the inertia of the electrons and holes are neglected as an approximation. Thus the theory cannot describe the plasma waves of electrons and holes. In many applications of semiconductors, plasma waves of mobile charges need to be considered and they can interact with elastic waves [2–4]. In this paper, we generalize [1] to include the inertia of the electron and hole fluids and allow the internal densities of the fluids to depend on the absolute temperature too in addition to their charge densities. The generalization is based on essentially the same multi-continuum model of several charged and interpenetrating continua in [1]. The generalization may be considered as straightforward but it is somewhat lengthy. Similar multi-continuum models have been used to derive equations for other elastic-electromagnetic interactions [5–15]. In general, the multi-continuum model may be viewed as a mixture of elastic-electromagnetic constituents [16]. The multi-continuum model is simpler than the fundamental model of charged particle [17] widely used in continuum electrodynamics and is sufficient for many purposes in applications.

## 2. Multi-Continuum Model

The model consist of six constituents of the lattice, bound charge, circulating current (or spin for simplicity), impurity, electron and hole continua [1]. The lattice, circulating current, and impurity continua are assumed to be moving together. The electron and hole fluids can flow through the lattice. Fields associated with the electron and hole fluids are indicated by superscripts $e$ and $h$. The bound-charge continuum can displace with respect to the lattice by a small displacement $\boldsymbol{\eta}$ for describing electric polarization. We will consider the inertia of the lattice (including the inertia of the impurity) and the inertia of the mobile charge fluids. The bound charge and spin continua are considered as massless. In the following, we present mainly the most basic equations and those that are affected by the mobile charge inertia. Please see [1] for more details about the notation and the equations that are not affected by the mobile charge inertia. However, [1] is in Gaussian units. Some of the equations can also be found in [15] in SI units. Let the lattice be moving with $\mathbf{y} = \mathbf{y}(\mathbf{X}, t)$ where $\mathbf{y}$ are the spatial coordinates of the material particle of the lattice who was initially at $\mathbf{X}$ in the reference state. We have, for the lattice charge density $\mu^l$, bound charge density $\mu^b$ and residual charge density $\mu^r$, the following relationship [1]:

$$\mu^l(\mathbf{y}) + \mu^b(\mathbf{y}+\boldsymbol{\eta}) = \mu^r(\mathbf{y}). \tag{2.1}$$

It is assumed that $\boldsymbol{\eta}$ satisfies [1]

$$\eta_{k,k} = 0. \tag{2.2}$$

The electric polarization $\mathbf{P}$ is given by [1]

$$\mathbf{P} = \mu^l(\mathbf{y})(-\boldsymbol{\eta}) = \mu^b(\mathbf{y}+\boldsymbol{\eta})\boldsymbol{\eta} \cong \mu^b(\mathbf{y})\boldsymbol{\eta}, \tag{2.3}$$

where only linear terns of $\boldsymbol{\eta}$ is kept. We denote the total charge density $\mu$ and current densities $\mathbf{J}$ by [1]

$$\mu = \mu^r + \mu^e + \mu^h + \mu^i, \tag{2.4}$$

$$\mathbf{J} = \mu^r \mathbf{v} + \mu^i \mathbf{v} + \mu^e \mathbf{v}^e + \mu^h \mathbf{v}^h. \tag{2.5}$$

$\mu^i$ is the impurity charge density. $\mathbf{v}$, $\mathbf{v}^e$ and $\mathbf{v}^h$ are the velocity fields of the lattice, electron and hole continua in the fixed reference frame. The electromagnetic body force, couple and power on a unit volume of the combined continuum at $\mathbf{y}$ including the corresponding bound-charge continuum at $\mathbf{y}+\boldsymbol{\eta}$ can be written as [1,15]

$$\begin{aligned}\mathbf{F}^{EM} &= \mathbf{P}\cdot\nabla\mathbf{E} + \mathbf{M}'\cdot(\mathbf{B}\nabla) + \mathbf{v}\times(\mathbf{P}\cdot\nabla\mathbf{B}) + \rho\dot{\boldsymbol{\pi}}\times\mathbf{B} \\ &\quad + \mu\mathbf{E} + \mathbf{J}\times\mathbf{B},\end{aligned} \tag{2.6}$$

where $\mathbf{M}'$ is the magnetization per unit volume in the co-moving or rest frame of the lattice continuum, $\boldsymbol{\pi}$ is the polarization per unit mass of the lattice continuum, and a superimposed dot represents the material time derivative $d/dt$ following the lattice continuum. The electromagnetic body force can also be written as [1,15]

$$F_j^{EM} = T_{ij,i}^{EM} - \frac{\partial G_j}{\partial t}, \tag{2.7}$$

where $\mathbf{G}$ is the electromagnetic momentum density [1,15]

$$\mathbf{G} = \varepsilon_0 \mathbf{E}\times\mathbf{B}, \quad G_j = \varepsilon_0 \varepsilon_{jkl} E_k B_l, \tag{2.8}$$

and $\mathbf{T}^{EM}$ is the electromagnetic stress tensor [1,15]

$$\begin{aligned}T_{ij}^{EM} &= P_i E_j' - B_i M_j' + \varepsilon_0 E_i E_j + \frac{1}{\mu_0} B_i B_j \\ &\quad - \frac{1}{2}\left(\varepsilon_0 E_k E_k + \frac{1}{\mu_0} B_k B_k - 2M_k' B_k\right)\delta_{ij}.\end{aligned} \tag{2.9}$$

The electromagnetic couple on a unit volume of the combined continuum is [1,15]

$$\mathbf{C}^{EM} = \mathbf{P}\times\mathbf{E}' + \mathbf{M}'\times\mathbf{B}, \tag{2.10}$$

where $\mathbf{E}'$ is the electric field in the rest frame of the lattice continuum. The electromagnetic body power on a unit volume of the combined continuum is [1,15]

$$\begin{aligned}W^{EM} &= (\mathbf{P}\cdot\nabla\mathbf{E})\cdot\mathbf{v} + \rho\mathbf{E}\cdot\dot{\boldsymbol{\pi}} - \mathbf{M}'\cdot\frac{\partial\mathbf{B}}{\partial t} + \mathbf{E}\cdot\mathbf{J} \\ &= \mathbf{F}^{EM}\cdot\mathbf{v} + \rho\mathbf{E}'\cdot\dot{\boldsymbol{\pi}} - \mathbf{M}'\cdot\dot{\mathbf{B}} + \mathbf{J}'\cdot\mathbf{E}',\end{aligned} \tag{2.11}$$

where

$$\mathbf{J}' = \mathbf{J} - \mu\mathbf{v} = \mu^e(\mathbf{v}^e - \mathbf{v}) + \mu^h(\mathbf{v}^h - \mathbf{v}). \tag{2.12}$$

## 3. Integral Balance Laws

Since different constituents have different velocities, we begin with the spatial description. Consider a fixed spatial region $v$ with a boundary surface $s$ whose outward unit normal is $\mathbf{n}$. The conservation of mass of the lattice, impurity, bound charge and spin continua together is [1]:

$$\frac{\partial}{\partial t}\int_v \rho\, dv = -\int_s \mathbf{n}\cdot\mathbf{v}\rho\, ds. \tag{3.1}$$

The conservation of residual, impurity, electron and hole charges are [1]:

$$\frac{\partial}{\partial t}\int_v \mu^r dv = -\int_s \mathbf{n}\cdot\mathbf{v}\mu^r ds, \tag{3.2}$$

$$\frac{\partial}{\partial t}\int_v \mu^i dv = -\int_s \mathbf{n}\cdot\mathbf{v}\mu^i ds, \tag{3.3}$$

$$\frac{\partial}{\partial t}\int_v \mu^e dv = -\int_s \mathbf{n}\cdot\mathbf{v}^e\mu^e ds, \tag{3.4}$$

$$\frac{\partial}{\partial t}\int_v \mu^h dv = -\int_s \mathbf{n}\cdot\mathbf{v}^h\mu^h ds, \tag{3.5}$$

where, for simplicity, the exchanges of charges among the constituents [1] are not considered here. We note that the conservation of mass for the electron and hole continua are not independent to their conservation of charge because of their fixed charge-to-mass ratios. The addition of Eqs. (3.2)–(3.5) yields [1]:

$$\frac{\partial}{\partial t}\int_v \mu\, dv = -\int_s \mathbf{n}\cdot\mathbf{J}\, ds. \tag{3.6}$$

The integral forms of Maxwell's equations are [1]:

$$\oint_l \mathbf{E}\cdot\mathbf{dl} = -\frac{\partial}{\partial t}\int_s \mathbf{n}\cdot\mathbf{B}\, ds, \tag{3.7}$$

$$\oint_l \mathbf{H}\cdot\mathbf{dl} = \frac{\partial}{\partial t}\int_s \mathbf{n}\cdot\mathbf{D}\, ds + \int_s \mathbf{n}\cdot\mathbf{J}\, ds, \tag{3.8}$$

$$\int_s \mathbf{n}\cdot\mathbf{D}\, ds = \mu, \tag{3.9}$$

$$\int_s \mathbf{n}\cdot\mathbf{B}\, ds = 0. \tag{3.10}$$

The linear momentum equation of the lattice, impurity, bound-charge and spin continua together is

$$\begin{aligned}\frac{\partial}{\partial t}\int_v \rho\mathbf{v}\, dv &= -\int_s (\mathbf{n}\cdot\mathbf{v})\rho\mathbf{v}\, dv + \int_s \mathbf{t}\, ds \\ &+ \int_v (\mathbf{P}\cdot\nabla\mathbf{E} + \mathbf{M}'\cdot(\mathbf{B}\nabla) + \mathbf{v}\times(\mathbf{P}\cdot\nabla\mathbf{B}) + \rho\dot{\boldsymbol{\pi}}\times\mathbf{B})dv \\ &+ \int_v [\rho\mathbf{f} + (\mu^r + \mu^i)\mathbf{E} + (\mu^r + \mu^i)\mathbf{v}\times\mathbf{B} - \mu^e\mathbf{E}^e - \mu^h\mathbf{E}^h]dv,\end{aligned} \tag{3.11}$$

where $\mathbf{t}$ is surface traction per unit area and $\mathbf{f}$ is body force per unit mass. $\mathbf{E}^e$ and $\mathbf{E}^h$ represent the interactions between the lattice and the mobile charges [1]. The electron and hole continua are assumed to be ideal fluids with pressure fields $p^e$ and $p^h$ [1], respectively. Their linear momentum equations assume the following form:

$$\frac{\partial}{\partial t}\int_v \rho^e \mathbf{v}^e dv = -\int_s (\mathbf{n}\cdot\mathbf{v}^e)\rho^e \mathbf{v}^e dv \\ + \int_s -p^e \mathbf{n} ds + \int_v [\rho^e \mathbf{f}^e + \mu^e(\mathbf{E}+\mathbf{v}^e\times\mathbf{B}) + \mu^e \mathbf{E}^e] dv, \quad (3.12)$$

$$\frac{\partial}{\partial t}\int_v \rho^h \mathbf{v}^h dv = -\int_s (\mathbf{n}\cdot\mathbf{v}^h)\rho^h \mathbf{v}^h dv \\ + \int_s -p^h \mathbf{n} ds + \int_v [\rho^h \mathbf{f}^h + \mu^h(\mathbf{E}+\mathbf{v}^h\times\mathbf{B}) + \mu^h \mathbf{E}^h] dv. \quad (3.13)$$

The addition of Eqs. (3.11)–(3.13) gives the following linear momentum equation for all constituents together:

$$\frac{\partial}{\partial t}\int_v (\rho\mathbf{v} + \rho^e\mathbf{v}^e + \rho^h\mathbf{v}^h) dv \\ = -\int_s [(\mathbf{n}\cdot\mathbf{v})\rho\mathbf{v} + (\mathbf{n}\cdot\mathbf{v}^e)\rho^e\mathbf{v}^e + (\mathbf{n}\cdot\mathbf{v}^h)\rho^h\mathbf{v}^h] dv \\ + \int_s (\mathbf{t} - p^e\mathbf{n} - p^h\mathbf{n}) ds + \int_v [\rho\mathbf{f} + \rho^e\mathbf{f}^e + \rho^h\mathbf{f}^h + \mathbf{F}^{EM}] dv. \quad (3.14)$$

The angular momentum equation of the combined continuum of all constituents is

$$\frac{\partial}{\partial t}\int_v \mathbf{y}\times(\rho\mathbf{v} + \rho^e\mathbf{v}^e + \rho^h\mathbf{v}^h) dv \\ = -\int_s \left[(\mathbf{n}\cdot\mathbf{v})\mathbf{y}\times\rho\mathbf{v} + (\mathbf{n}\cdot\mathbf{v}^e)\mathbf{y}\times\rho^e\mathbf{v}^e + (\mathbf{n}\cdot\mathbf{v}^h)\mathbf{y}\times\rho^h\mathbf{v}^h\right] ds \\ + \int_s \mathbf{y}\times(\mathbf{t} - p^e\mathbf{n} - p^h\mathbf{n}) ds \\ + \int_v \left[\mathbf{y}\times(\rho\mathbf{f} + \rho^e\mathbf{f}^e + \rho^h\mathbf{f}^h + \mathbf{F}^{EM}) + \mathbf{C}^{EM}\right] dv. \quad (3.15)$$

The energy equation of the combined continuum of all constituents can be written as

$$\frac{\partial}{\partial t}\int_v \left(\frac{1}{2}\rho\mathbf{v}\cdot\mathbf{v} + \frac{1}{2}\rho^e\mathbf{v}^e\cdot\mathbf{v}^e + \frac{1}{2}\rho^h\mathbf{v}^h\cdot\mathbf{v}^h\right) dv + \frac{\partial}{\partial t}\int_v (\rho\varepsilon + \rho^e\varepsilon^e + \rho^h\varepsilon^h) dv \\ = \int_s (\mathbf{t}\cdot\mathbf{v} - p^e\mathbf{n}\cdot\mathbf{v}^e - p^h\mathbf{n}\cdot\mathbf{v}^h - \mathbf{n}\cdot\mathbf{q}) ds \\ - \int_s \mathbf{n}\cdot\left(\mathbf{v}\frac{1}{2}\rho\mathbf{v}\cdot\mathbf{v} + \mathbf{v}^e\frac{1}{2}\rho^e\mathbf{v}^e\cdot\mathbf{v}^e + \mathbf{v}^h\frac{1}{2}\rho^h\mathbf{v}^h\cdot\mathbf{v}^h\right) ds \\ - \int_s \mathbf{n}\cdot(\mathbf{v}\rho\varepsilon + \mathbf{v}^e\rho^e\varepsilon^e + \mathbf{v}^h\rho^h\varepsilon^h) ds \\ + \int_v (\rho\mathbf{f}\cdot\mathbf{v} + \rho^e\mathbf{f}^e\cdot\mathbf{v}^e + \rho^h\mathbf{f}^h\cdot\mathbf{v}^h + \rho r) dv + \int_v W^{EM} dv, \quad (3.16)$$

where $\varepsilon$ is internal energy density, $\mathbf{q}$ the heat flux vector, and $r$ the body heat source. Let the entropy density per unit mass be $\eta$. Then the entropy inequality of all constituents together is

$$\frac{\partial}{\partial t}\int_v (\rho\eta + \rho^e\eta^e + \rho^h\eta^h)dv$$
$$\geq -\int_s (\rho\eta\mathbf{v}\cdot\mathbf{n} + \rho^e\eta^e\mathbf{v}^e\cdot\mathbf{n} + \rho^h\eta^h\mathbf{v}^h\cdot\mathbf{n})ds + \int_v \frac{\rho r}{\theta}dv - \int_s \frac{\mathbf{q}\cdot\mathbf{n}}{\theta}ds. \tag{3.17}$$

## 4. Differential Balance Laws

The differential forms of the balance laws corresponding to Eqs. (3.1–3.13) can be found as

$$\frac{\partial \rho}{\partial t} + \nabla\cdot(\rho\mathbf{v}) = 0, \tag{4.1}$$

$$\frac{\partial \mu^r}{\partial t} + \nabla\cdot(\mu^r\mathbf{v}) = 0, \tag{4.2}$$

$$\frac{\partial \mu^i}{\partial t} + \nabla\cdot(\mu^i\mathbf{v}) = 0, \tag{4.3}$$

$$\frac{\partial \mu^e}{\partial t} + \nabla\cdot(\mu^e\mathbf{v}^e) = 0, \tag{4.4}$$

$$\frac{\partial \mu^h}{\partial t} + \nabla\cdot(\mu^h\mathbf{v}^h) = 0, \tag{4.5}$$

$$\frac{\partial \mu}{\partial t} + \nabla\cdot\mathbf{J} = 0, \tag{4.6}$$

$$\nabla\times\mathbf{E} = -\frac{\partial \mathbf{B}}{\partial t}, \tag{4.7}$$

$$\nabla\times\mathbf{H} = \frac{\partial \mathbf{D}}{\partial t} + \mathbf{J}, \tag{4.8}$$

$$\nabla\cdot\mathbf{D} = \mu, \tag{4.9}$$

$$\nabla\cdot\mathbf{B} = 0, \tag{4.10}$$

$$\nabla\cdot\boldsymbol{\tau} + \rho\mathbf{f} + \mathbf{P}\cdot\nabla\mathbf{E} + \mathbf{M}'\cdot(\mathbf{B}\nabla) + \mathbf{v}\times(\mathbf{P}\cdot\nabla\mathbf{B}) + \rho\dot{\boldsymbol{\pi}}\times\mathbf{B}$$
$$+ (\mu^r + \mu^i)\mathbf{E} + (\mu^r + \mu^i)\mathbf{v}\times\mathbf{B} - \mu^e\mathbf{E}^e - \mu^h\mathbf{E}^h = \rho\frac{d\mathbf{v}}{dt}, \tag{4.11}$$

$$-\nabla p^e + \rho^e\mathbf{f}^e + \mu^e\left(\mathbf{E} + \mathbf{v}^e\times\mathbf{B} + \mathbf{E}^e\right) = \rho^e\frac{d^e\mathbf{v}^e}{dt}, \tag{4.12}$$

$$-\nabla p^h + \rho^h\mathbf{f}^h + \mu^h\left(\mathbf{E} + \mathbf{v}^h\times\mathbf{B} + \mathbf{E}^h\right) = \rho^h\frac{d^h\mathbf{v}^h}{dt}, \tag{4.13}$$

where $\boldsymbol{\tau}$ is the stress tensor in the lattice continuum. $d/dt$, $d^e/dt$ and $d^h/dt$ are material time derivatives following the lattice, electron fluid, and hole fluid, respectively. In the derivation of the above linear momentum equations, the continuity equations from conservation of mass have been used. Adding Eqs. (4.11–4.13), we obtain

$$\nabla \cdot \boldsymbol{\tau} - \nabla p^e - \nabla p^h + \rho \mathbf{f} + \rho^e \mathbf{f}^e + \rho^h \mathbf{f}^h + \mathbf{F}^{EM}$$
$$= \rho \frac{d\mathbf{v}}{dt} + \rho^e \frac{d^e \mathbf{v}^e}{dt} + \rho^h \frac{d^h \mathbf{v}^h}{dt}, \tag{4.14}$$

which corresponds to Eq. (3.14). The angular momentum equation, energy equation and entropy inequality corresponding to Eqs. (3.15–3.17) are

$$\varepsilon_{kij} \tau_{ij} + C_k^{EM} = 0. \tag{4.15}$$

$$\rho \frac{d\varepsilon}{dt} + \rho^e \frac{d^e \varepsilon^e}{dt} + \rho^h \frac{d^h \varepsilon^h}{dt} - \frac{p^e}{\rho^e} \frac{d^e \rho^e}{dt} - \frac{p^h}{\rho^h} \frac{d^h \rho^h}{dt}$$
$$= \tau_{ij} v_{j,i} + \rho E'_i \frac{d\pi_i}{dt} - M'_i \frac{dB_i}{dt} \tag{4.16}$$
$$- \mu^e \mathbf{E}^e \cdot (\mathbf{v}^e - \mathbf{v}) - \mu^h \mathbf{E}^h \cdot (\mathbf{v}^h - \mathbf{v}) + \rho r - q_{i,i},$$

$$\rho \frac{d\eta}{dt} + \rho^e \frac{d^e \eta^e}{dt} + \rho^h \frac{d^h \eta^h}{dt} \geq \frac{\rho r}{\theta} - \left(\frac{q_i}{\theta}\right)_{,i}. \tag{4.17}$$

In the derivation of Eqs. (4.15–4.17) the continuity and linear momentum equations have been used.

## 5. Constitutive Relations

Eliminating $r$ from Eqs. (4.16) and (4.17), we obtain the Clausius–Duhem inequality as

$$\rho \left( \theta \frac{d\eta}{dt} - \frac{d\varepsilon}{dt} \right) + \rho^e \left( \theta \frac{d^e \eta^e}{dt} - \frac{d^e \varepsilon^e}{dt} \right) + \rho^h \theta \left( \frac{d^h \eta^h}{dt} - \frac{d^h \varepsilon^h}{dt} \right)$$
$$+ \frac{p^e}{\rho^e} \frac{d^e \rho^e}{dt} + \frac{p^h}{\rho^h} \frac{d^h \rho^h}{dt} + \tau_{ij} v_{j,i} + \rho E'_i \frac{d\pi_i}{dt} - M'_i \frac{dB_i}{dt} \tag{5.1}$$
$$- \mu^e \mathbf{E}^e \cdot (\mathbf{v}^e - \mathbf{v}) - \mu^h \mathbf{E}^h \cdot (\mathbf{v}^h - \mathbf{v}) - \frac{q_i \theta_{,i}}{\theta} \geq 0.$$

The following free energy densities for different constituents can be introduced through the corresponding Legendre transforms:

$$F = \varepsilon - \mathbf{E}' \cdot \boldsymbol{\pi} - \eta \theta,$$
$$F^e = \varepsilon^e - \theta \eta^e, \quad F^h = \varepsilon^h - \theta \eta^h. \tag{5.2}$$

Then the energy equation and Clausius–Duhem inequality take the following form:

$$\rho \left( \frac{dF}{dt} + \theta \frac{d\eta}{dt} + \eta \frac{d\theta}{dt} \right) + \rho^e \left( \frac{d^e F^e}{dt} + \theta \frac{d^e \eta^e}{dt} + \eta^e \frac{d^e \theta}{dt} \right)$$
$$+ \rho^h \left( \frac{d^h F^h}{dt} + \theta \frac{d^h \eta^h}{dt} + \eta^h \frac{d^h \theta}{dt} \right) - \frac{p^e}{\rho^e} \frac{d^e \rho^e}{dt} - \frac{p^h}{\rho^h} \frac{d^h \rho^h}{dt} \tag{5.3}$$
$$= \tau_{ij} v_{j,i} - P_i \frac{dE'_i}{dt} - M'_i \frac{dB_i}{dt} - \mu^e \mathbf{E}^e \cdot (\mathbf{v}^e - \mathbf{v}) - \mu^h \mathbf{E}^h \cdot (\mathbf{v}^h - \mathbf{v}) + \rho r - q_{i,i},$$

$$-\rho\left(\frac{dF}{dt}+\eta\frac{d\theta}{dt}\right)-\rho^e\left(\frac{d^eF^e}{dt}+\eta^e\frac{d^e\theta}{dt}\right)-\rho^h\left(\frac{d^hF^h}{dt}+\eta^h\frac{d^h\theta}{dt}\right)$$
$$+\frac{p^e}{\rho^e}\frac{d^e\rho^e}{dt}+\frac{p^h}{\rho^h}\frac{d^h\rho^h}{dt}+\tau_{ij}v_{j,i}-P_i\frac{dE'_i}{dt}-M'_i\frac{dB_i}{dt} \quad (5.4)$$
$$-\mu^e\mathbf{E}^e\cdot(\mathbf{v}^e-\mathbf{v})-\mu^h\mathbf{E}^h\cdot(\mathbf{v}^h-\mathbf{v})-\frac{q_i\theta_{,i}}{\theta}\geq 0.$$

We break the stress, polarization and magnetization into recoverable and dissipative parts as

$$\boldsymbol{\tau}=\boldsymbol{\tau}^R+\boldsymbol{\tau}^D,\quad \mathbf{P}=\mathbf{P}^R+\mathbf{P}^D,\quad \mathbf{M}'=\mathbf{M}'^R+\mathbf{M}'^D. \quad (5.5)$$

The recoverable parts satisfy

$$\rho\left(\frac{dF}{dt}+\eta\frac{d\theta}{dt}\right)+\rho^e\left(\frac{d^eF^e}{dt}+\eta^e\frac{d^e\theta}{dt}\right)+\rho^f\left(\frac{d^fF^f}{dt}+\eta^f\frac{d^f\theta}{dt}\right)$$
$$-\frac{p^e}{\rho^e}\frac{d^e\rho^e}{dt}-\frac{p^h}{\rho^h}\frac{d^h\rho^h}{dt}=\tau^R_{ij}v_{j,i}-P^R_i\frac{dE'_i}{dt}-M'^R_i\frac{dB_i}{dt}. \quad (5.6)$$

Then the energy equation and Clausius–Duhem inequality reduce to

$$\rho\theta\frac{d\eta}{dt}+\rho^e\theta\frac{d^e\eta^e}{dt}+\rho^h\theta\frac{d^h\eta^h}{dt}=\tau^D_{ij}v_{j,i}-P^D_j\frac{dE'_j}{dt}-M'^D_j\frac{dB_j}{dt}$$
$$-\mu^e\mathbf{E}^e\cdot(\mathbf{v}^e-\mathbf{v})-\mu^h\mathbf{E}^h\cdot(\mathbf{v}^h-\mathbf{v})+\rho r-q_{i,i}, \quad (5.7)$$

$$\tau^D_{ij}v_{j,i}-P^D_j\frac{dE'_j}{dt}-M'^D_j\frac{dB_j}{dt}-\mu^e\mathbf{E}^e\cdot(\mathbf{v}^e-\mathbf{v})-\mu^h\mathbf{E}^h\cdot(\mathbf{v}^h-\mathbf{v})-\frac{q_i\theta_{,i}}{\theta}\geq 0. \quad (5.8)$$

Equation (5.7) is the dissipation equation. Equation (5.8) imposes restrictions on the dissipative parts of the constitutive relations.

For the recoverable fields satisfying Eq. (5.6), consider

$$F=F(v_{j,i};\mathbf{E}';\mathbf{B};\theta),\quad F^e=F^e\left[(\rho^e)^{-1};\theta\right],\quad F^h=F^h\left[(\rho^h)^{-1};\theta\right]. \quad (5.9)$$

Since

$$v_{j,i}=X_{M,i}\frac{d}{dt}(y_{j,M}), \quad (5.10)$$

we can write Eq. (5.6) as

$$\left[X_{M,i}\tau^R_{ij}-\rho\frac{\partial F}{\partial(y_{j,M})}\right]\frac{d}{dt}(y_{j,M})-\left(P^R_i+\rho\frac{\partial F}{\partial E'_i}\right)\frac{dE'_i}{dt}-\left(M'^R_i+\rho\frac{\partial F}{\partial B_i}\right)\frac{dB_i}{dt}$$
$$-\rho\left(\eta+\frac{\partial F}{\partial\theta}\right)\frac{d\theta}{dt}-\rho^e\left(\eta^e+\frac{\partial F^e}{\partial\theta}\right)\frac{d^e\theta}{dt}-\rho^h\left(\eta^h+\frac{\partial F^h}{\partial\theta}\right)\frac{d^h\theta}{dt} \quad (5.11)$$
$$+\frac{1}{\rho^e}\left(p^e+\frac{\partial F^e}{\partial(\rho^e)^{-1}}\right)\frac{d^e\rho^e}{dt}+\frac{1}{\rho^h}\left(p^h+\frac{\partial F^h}{\partial(\rho^h)^{-1}}\right)\frac{d^h\rho^h}{dt}=0,$$

which implies the following recoverable constitutive relations:

$$\tau_{ij}^R = \rho y_{i,M} \frac{\partial F}{\partial (y_{j,M})}, \quad P_i^R = -\rho \frac{\partial F}{\partial E_i'}, \quad M_i'^R = -\rho \frac{\partial F}{\partial B_i},$$

$$\eta = -\frac{\partial F}{\partial \theta}, \quad \eta^e = -\frac{\partial F^e}{\partial \theta}, \quad \eta^h = -\frac{\partial F^h}{\partial \theta}, \tag{5.12}$$

$$p^e = -\frac{\partial F^e}{\partial (\rho^e)^{-1}}, \quad p^h = -\frac{\partial F^h}{\partial (\rho^h)^{-1}}.$$

For rotational invariance (0bjectivity) [1], $F$ can be reduced to a function of the following inner products and $\theta$ [1]:

$$C_{KL} = y_{i,K} y_{i,L}, \quad \mathcal{E}_L = y_{i,L} E_i', \quad Z_L = y_{i,L} B_i. \tag{5.13}$$

We will use the strain tensor $E_{KL}$ instead of the deformation tensor $C_{KL}$. Therefore, we take

$$F = F(E_{KL}; \mathcal{E}_K; Z_K; \theta). \tag{5.14}$$

Then the first four of the constitutive relations in Eq. (5.8) become

$$\tau_{ij}^R = \rho y_{i,M} \frac{\partial F}{\partial E_{ML}} y_{j,L} + \rho y_{i,M} \frac{\partial F}{\partial \mathcal{E}_M} E_j' + \rho y_{i,M} \frac{\partial F}{\partial Z_M} B_j,$$

$$P_i^R = -\rho y_{i,L} \frac{\partial F}{\partial \mathcal{E}_L}, \quad M_i'^R = -\rho y_{i,L} \frac{\partial F}{\partial Z_L}, \quad \eta = -\frac{\partial F}{\partial \theta}. \tag{5.15}$$

The dissipative parts of the constitutive relations may assume the following forms in general:

$$\boldsymbol{\tau}^D = \boldsymbol{\tau}^D(y_{j,M}, \mathbf{E}', \mathbf{B}, \mu^e, \mu^h, \mathbf{E}^e, \mathbf{E}^h, \theta),$$

$$\mathbf{P}^D = \mathbf{P}^D(y_{j,M}, \mathbf{E}', \mathbf{B}, \mu^e, \mu^h, \mathbf{E}^e, \mathbf{E}^h, \theta), \tag{5.16}$$

$$\mathbf{M}'^D = \mathbf{M}'^D(y_{j,M}, \mathbf{E}', \mathbf{B}, \mu^e, \mu^h, \mathbf{E}^e, \mathbf{E}^h, \theta).$$

Similarly, the heat flux (Fourier's law) may be [1]

$$\mathbf{q} = \mathbf{q}(y_{j,M}, \mathbf{E}', \mathbf{B}, \mu^e, \mu^h, \mathbf{E}^e, \mathbf{E}^h, \theta, \theta_{,k}). \tag{5.17}$$

For the interactions among the constituents, while constitutive relations may be given in terms of the interactions represented by $\mathbf{E}^e$ and $\mathbf{E}^h$, they are usually given for $\mathbf{v}^e$–$\mathbf{v}$ and $\mathbf{v}^h$–$\mathbf{v}$ for conduction (Ohm's law) [1]:

$$\mathbf{v}^e - \mathbf{v} = \mathbf{V}^e(y_{j,M}, \mathbf{E}', \mathbf{B}, \mu^e, \mathbf{E}^e, \theta, \theta_{,k}),$$

$$\mathbf{v}^h - \mathbf{v} = \mathbf{V}^h(y_{j,M}, \mathbf{E}', \mathbf{B}, \mu^h, \mathbf{E}^h, \theta, \theta_{,k}). \tag{5.18}$$

The dissipative constitutive relations are restricted by the Clausius–Duhem inequality in Eq. (7.4) and need to satisfy the requirements of rotational invariance [1].

## 6. Conclusions

In summary, the basic unknown fields are $\rho$, $\rho^e$, $\rho^h$, $\mathbf{E}$, $\mathbf{B}$, $\mathbf{H}$, $\mathbf{D}$, $\mathbf{v}$, $\mathbf{v}^e$, $\mathbf{v}^h$ and the temperature field $\theta$. The basic field equations are the continuity equations in Eqs. (4.1), (4.4) and (4.5), Maxwell's equations, the linear momentum equations in Eqs. (4.11)–(4.13), and the dissipation equation in Eq. (5.7). There are other fields but they are related to the above ones by constitutive relations, charge-to-mass ratios, and kinematic relations, etc. On the boundary surface of a finite body, the tractions or velocities of the constituents, the boundary conditions associated with Maxwell's equations [1], and the temperature or the normal heat flux may be prescribed. These equations have mobile charge inertia and can describe plasma waves. They can be further generalized to include the exchange of mass, charge, energy and momenta among the constituents.